\documentstyle[12pt,epsf,eqalign]{article}

\newcommand{\etal}{{\it et~al., }}
\newcommand{\etl}{{\it et~al.}}

\newcommand{\estim}[2]{{}\>{#1}\>{#2}\'\\}

\begin{document}

\title{RELATIVISTIC BINARY MERGING RATES}
\author{V.~M.~Lipunov\\
 Sternberg Astronomical Institute, Moscow}

\date{}
\maketitle

\vskip 1cm
\begin{abstract}
The Invited Review on Joint Discussion "High Energy Transients" on
XXIIIrd General Assembly of IAU, Kioto, 1997 is presented.
The simulated rates of neutron star + neutron star, neutron star + 
black hole and black hole + black hole mergings are
considered in relation with the problem of GRBursts origin
and gravitational waves detection.
\end{abstract}

\vskip 2cm
\section*{Observations}

After the outstanding experiment BeppoSAX (Costa \etal 1997, IAUC~6572)
and discovery of optical afterglow phenomenon in GRB~970228 (Groot \etal
1997, IAUC~6584; Sahu \etal
1997, IAUC~6606) and discovery of spectral lines in GRB~970508 ($z=0.835$)
(Metzger \etal
1997, IAUC~6655) we know that in the Universe there are real sources
with luminosities more than $10^{50}\, \mbox{erg}/\mbox{s}$.
The mergings of binary relativistic stars are the most powerful
high energy transients in the Universe: the released power is
of order of the Planckian one $c^5/G\sim 10^{58}$~erg/s.

The mergings of relativistic binaries may underlie the origin of
cosmic gamma--ray bursts (GRB)
(Blinnikov \etal 1984; Pazcy\'nski, 1986; Meszaros and Rees, 1992).

There are three types of merging reactions (``M--reaction") of
relativistic stars:
$$\eqalign{
 NS + NS &\rightarrow GWB + \nu B + GRB(?) + NS/BH\cr
         NS + BH &\rightarrow GWB + BH + \nu B + GRB(?) \cr
              BH + BH &\rightarrow GWB + BH \cr
}
$$

We see two branch of theoretical research:
\begin{itemize}
\item[a)] physical investigation of M--reactions (Mergingology:
 fairball formation, numerical relativity, hydrodynamics. The pulsar
mechanism also can act (Lipunov \& Panchenko 1996b), (Lipunova 1997).
\item[b)] astrophysical calculation of ``crossection" or
        ``probability" of the M--reaction in our Universe
        (Population synthesis).
\end{itemize}

 In a few years several initial ground--based laser interferometers
aimed at searching for gravitational waves (GW) will start to  work
(LIGO~(Abramovici \etl 1992), VIRGO (Ciufolini 1992), GEO--600~(Schutz
1996),\break
TAMA-300~(http://tamago.mtk.nao.ac.jp/)), so at present time the question:
what kind of events and how frequently will
the interferometer register? --- is very important.
Undoubtedly, the most reliable GW sources are the merging
compact binary stars --- double neutron stars (NS)
and black holes (BH) of different stellar masses.

\vskip 0.2cm
The current observational data:
\begin{enumerate}
\item A few binary radiopulsars are known to have the secondary NS
component.

\item Three of these binary pulsars must coalesce
due to the orbital angular momentum removal by GW
in a time scale shorter than the age of the Universe (the
Hubble time $t_H\simeq15\cdot 10^9$~yrs).

\item No binary pulsars with BH is known yet (although
from evolutionary considerations one may expect one such
object to be formed in the Galaxy per about 1000 single
pulsars, (Lipunov \etl 1994))

\item No binary BH has been found so far.

\item In contrast, 10 BH candidates are already known in
X-ray binary systems with normal companions (Cherepashchuk 1996).
Note that the mean BH mass
in these systems is
$<\mbox{M}_{\mbox{BH}}>\simeq 8.5~ \mbox{M}_{\odot}\,$,
i.e. BH formed in stellar evolution
are notably more massive than NS (with the typical mass
$1.4$M$_{\odot}$).

\end{enumerate}

\section*{Population synthesis: key parameters}\label{Scenario}

\subsection*{Binary NS Merging Rate Estimates}

At present time, it is possible to estimate binary NS merging rate in two ways:
using the binary radiopulsar statistics observed and making
various computations of binary stellar evolution (Population Synthesis).

\bigskip
\centerline{`Observational'' estimates}
\begin{tabbing}
\hspace{2.5cm}\=\hspace{11cm}\=\kill
\estim{(Phinney 1991)}{$1/10^6$~yr} 
\estim{(Narayan \etl 1991)}{$1/10^6$~yr} 
\estim{(Curran \& Lorimer 1995)}{$3/10^6$~yr}  
\estim{(van den Heuvel \& Lorimer 1996)}{$8/10^6$~yr}  
\estim{``Bailes limit''(Bailes 1996}{$<1/10^5$~yr}
\end{tabbing}
\centerline{`Theoretical'' estimates}
\begin{tabbing}
\hspace{2.5cm}\=\hspace{11cm}\=\kill
\estim{(Clark \etl 1979)}{$1/10^4$--$1/10^6$~yr}
\estim{(Lipunov \etl 1987)}{$1/10^4$~yr}
\estim{(Hils \etl 1991)}{$1/10^4$~yr}
\estim{(Tutukov \& Yungelson 1993)}{$3/10^4$--$1/10^4$~yr}
\estim{(Lipunov \etl 1995a)}{$<3/10^4$~yr}
\estim{(Portegies Zwart \& Spreeuw 1996)}{$3/10^5$~yr}
\estim{(Lipunov \etl 1997a)}{$3/10^4$--$3/10^5$~yr}
\end{tabbing}

\bigskip

We emphasize that although theoretical merging rates
are systematically higher than observational ones,
both estimates do not contradict each other.
The main
argument is that the first (observational) estimates
of binary NS merging rate are based on the statistics of binary
systems, in which only one of the components shines as radiopulsar,
which is not at all the {\it necessary} conditions for merging to occure
(Lipunov et al.1997a).

To calculate binary evolution, we have used
the population synthesis method (the Scenario Machine code),
which is in fact a version of Monte--Carlo calculations.
The most important (and practically unique) parameter changing the
galactic binary NS merging rate is the distribution of an additional
(kick) velocity imparted to NS at birth. 

The kick velocity distribution.
widely accepted now, is derived from the analysis of spatial
velocities of single radiopulsars (Lyne \& Lorimer, 1994).
One can approximate this 3--dimensional distribution
as
$$
f_{LL}(x)dx\propto x^{0.19}(1+x^{6.72})^{-1/2}dx
$$
where $x=w/w_o$ and the characteristic velocity
$w_o$ is a parameter in our calculations.
The Lyne \& Lorimer (1994)
pulsar transverse velocity distribution corresponds to
$w_o=400$~km/s. But statistic of binary PSRs gives { Kornilov and Lipunov (1984):}
$$
 \hbox{Mean Kick}= 75 \mbox{--} 100~ \mbox{km}/\mbox{s} \hbox{ -- for
Delta--function distribution}
$$
  Lipunov, Postnov \& Prokhorov (1996a, 1997a):
$$
   \hbox{Mean Kick} = 100\mbox{--}200~ \mbox{km}/\mbox{s} \hbox
{ -- for Maxwellian or Lyne \& Lorimer distribution}
$$

 New space velocity distribution for Radio--Pulsars ({ Hansen \&
Phinney, 1997}):
$$
      \hbox{Maxwellian} + \hbox{mean velocity} =
250\mbox{--}300~\mbox{km}/\mbox{s}~.
$$

\subsection*{Binary BH merging rate}
In contrast, for BH,  two additional parameters appear.
First of them is a threshold main sequence stellar mass
$M_{cr}$ for the star to collapse into a BH after
its nuclear evolution has ended.
This parameter is still poorly
determined and varies in a wide range:
e.g., according to
(van den Heuvel \& Habets 1984), $M_{cr}= 40$--80M$_{\odot}$;
(Tsujimot \etl 1997) give 40--60M$_\odot$; (Portegies Zwart \& Spreeuw 1996) 
derive $>$20M$_\odot$.

The second parameter is the fraction of the presupernova mass,
$k_{{bh}}$, collapsing into BH. This parameter is fully unknown, so
we varied it from 0.1 to 1 in our calculations.

\section*{Detection rate of binary compact star merging}\label{DetectionRate}

Under the assumptions made above, we can calculate
the binary merging rate $R$ in the Galaxy. The results are
presented in Fig.~\ref{f:R(w)}.
After having found the merging rate $R$ in a typical galaxy,
we need to go over the event rate $D$ at the detector.
Applying the optimal filtering technique (Thorne 1987), the
signal-to-noise ratio $S/N$ at the spiral-in stage is
$$ 
    \frac{S}{N} \propto \frac{M_{Óh}^{5/6}}{d} ~.
$$
Here $M_{ch}= (M_1M_2)^{3/5}(M_1+M_2)^{2/5}$ is ``chorp''--mass of the
binary system.
This means that for a given $S/N$ our detector
can register more massive BH from larger distances
than NS. The volume within which BH or NS is to be detected
should be proportional to $M_{ch}^{15/6}\,$. Then the ratio
of detection rates of BH and NS can be written as (Fig.2):
$$
    \frac{D_{BH}}{D_{NS}} = \frac{R_{BH}}{R_{NS}}
           \left( \frac{M_{BH}}{M_{NS}} \right)^{15/6}\,.
$$

\section*{Gamma--Ray Bursts}\label{GRB}
Using the dependence on time of compact binary merging rate
for "elliptical" galaxy (Lipunov \etl 1995b)
and assuming the cosmological origin of GRBs as products
of binary NS/NS coalescences, we can compute the theoretical
$\log~N$--$\log~S$ curve.

  Recently, Lipunov, Postnov and  Prokhorov (1997c)
estimated the redshift of GRB~970228 and GRB~970508 using the
mean statistical properties of observed GRBs.  They assume the
cosmological origin of GRBs as standard--candle binary neutron star
mergers.

Same result was obtained independently by Totany (1997).
Recent progress of observations of high redshift galaxies, however,
gives more detailed information on the cosmic star formation history
(Lilly \etal 1996; Madau \etal 1996). The Canada--France Redshift
Survey (CFRS) revealed a remarkable evolution of 2800 {\AA} luminosity
density, that is considered to be a star formation
indicator, as ${\cal L}_{2800} \propto
(1+z)^{3.9 \pm 0.75}$ to $z \sim 1$ (for $\Omega_0 = 1$, Lilly \etal 1996).
The constant SFR approximation in spiral galaxies is therefore no longer
justified even at $z < 1$.

 The redshift of GRB~970508 is apparentely
about 2, just below the upper limit that is recently determined, and
the absorption system at $z = 0.835$ seems not to be the site of the GRB.

\section*{Conclusion}

{\bf 1.} We estimate {\bf NS + NS} merging rate as follows:

\begin{tabbing}
\hspace{0.2cm}\=\hspace{14.6cm}\=\kill
\estim{1/10 $^4$~yrs}{per Galaxy}
\estim{{ 1/yr}}{for GEO-600, VIRGO, TAMA-300, LIGO-type detector ($h >
10^{-21}$)}
\estim{ 1/minute}{per Universe}
\end{tabbing}
\par\noindent
{\bf 2.} {\bf BH + BH} merging rate:\\
First LIGO--type interferometer events shell give us the 
simultaneous discovery of GRAVITATIONAL WAVES and BLACK HOLES
(Lipunov \etl 1997d), as the
expected detection rate for  BH+BH merging is
\begin{tabbing}
\hspace{0.2cm}\=\hspace{14.6cm}\=\kill
\estim{{\bf 10--100/yr }}{for LIGO--type detector ($h > 10^{-21}$)}
\end{tabbing}
\par\noindent
{\bf 3.} GRB mystery: assuming the relativistic binary merging as the origin
of the GRBs we obtain:

-- that $\log N$--$ \log S$ is fine;

 -- reasonable estimates of redshifts for February and May Beppo--Sax GRBs;

--   (NS $+$ NS)  ~ needs collimation (several degree);

 --    (NS $+$ BH)  ~ needs no anisotropy.

{

\begin{figure}[p]
\epsfxsize=0.9\hsize
\epsfbox{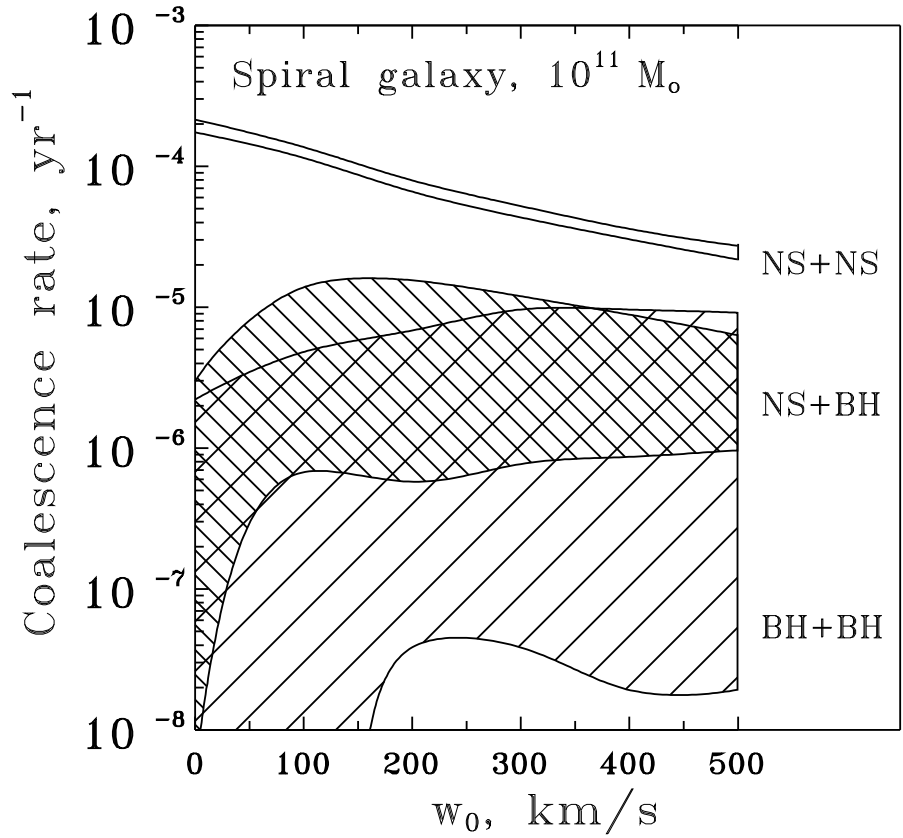}
\hss
\caption{{\bf  Lipunov, Postnov \& Prokhorov, 1997a} The dependence of different
compact binary systems coalescence rates on the characteristic
kick velocity $w_0$ in a spiral galaxy with $10^{11}$M${_\odot}$.}
\label{f:R(w)}
\end{figure}

\begin{figure}[p]
\hbox to\textwidth{\hss
\epsfxsize=0.8\hsize
\epsfbox{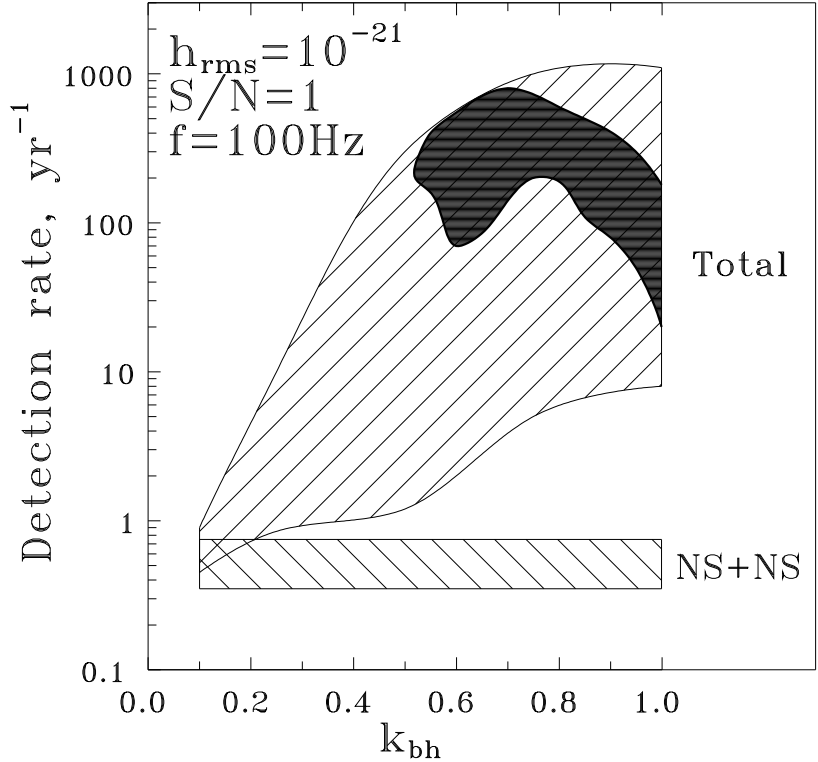}
\hss
}
\caption{{\bf Lipunov, Postnov \& Prokhorov, 1997b} The total merging rate of NS+NS, NS+BH, and BH+BH binaries
which would be detected by a laser interferometer with $h_{rms}=10^{-21}$
as a function of $k_{bh}$ for
Lyne--Lorimer kick velocity distribution with $w_0=200$--400~km/s and BH
progenitor's masses $M_{\ast}=15$--50M$_\odot$, for different scenarios of
binary star evolution. NS+NS mergings are shown separately.
In all cases BH+BH mergings contribute more than
$80\%$ to the total rate. The filled ``Loch--Ness--monster--head''--like
region corresponds to BH formation parameters
$M_{\ast}>18$M$_\odot$ and $k_{bh} > 0.5$.
}\label{f:sheja}
\end{figure}


\begin{figure}
\hbox to\textwidth{\hss
\epsfxsize=0.9\hsize
\epsfbox{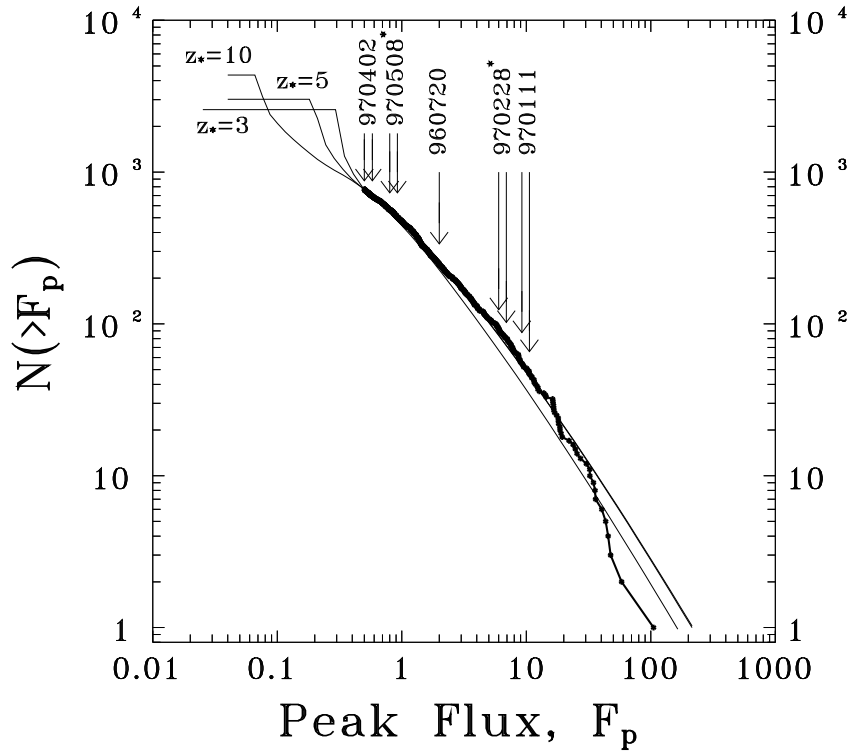}
\hss
}
\caption{{\bf Lipunov, Postnov \protect\& Prokhorov, 1997c}
The $\log N$--$\log F_{peak}$ distribution of 3B BATSE GRBs from
256-ms 1--3 (50--300 keV) channels fitted with the cosmological model
distributions in a flat, $\Omega=1$, Universe with a cosmological term
$\Omega_\Lambda=0.7$ assuming gamma-ray photon power law $s=-1.1$. The
locations of Beppo--SAX GRBs are shown.
GRB970228 and GRB970508 are marked with asterisks.}
\label{fig1}
\end{figure}

\begin{figure}
\hbox to\textwidth{\hss
\epsfxsize=0.9\hsize
\epsfbox{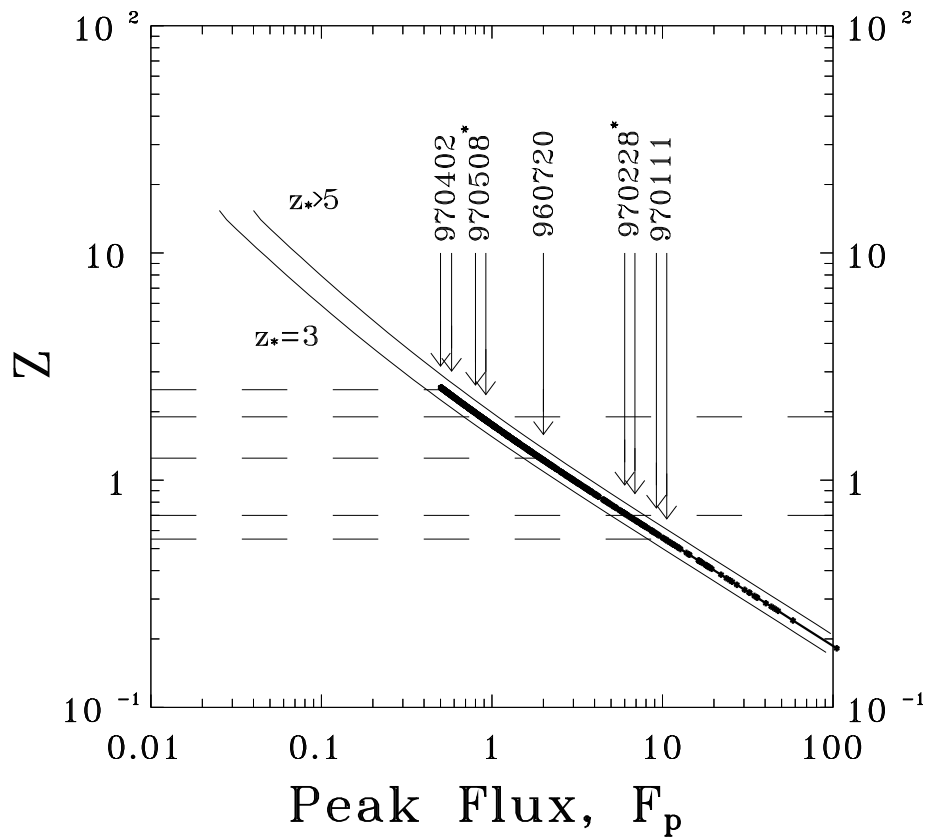}
\hss
}
\caption{{\bf Lipunov, Postnov \& Prokhorov, 1997, Astro-ph/9703181}
The redshift -- peak flux dependence
in the cosmologocal models assumed for different
$z_*$ and $s=-1.1$. 3B BATSE catalog data are also plotted.}
\label{fig2}
\end{figure}

}
\end{document}